\documentstyle[sprocl]{article}

\begin{document}

\title{ON FRACTIONAL SUPERSYMMETRIC QUANTUM MECHANICS: 
THE FRACTIONAL SUPERSYMMETRIC OSCILLATOR}

\author{ M. DAOUD }

\address{D\'epartement de Physique\\
Facult\'e des Sciences\\
Universit\'e Ibnou Zohr\\
B.P. 28/S, Agadir, Morocco}

\author{ M.R. KIBLER }

\address{Institut de Physique Nucl\'eaire de Lyon\\
IN2P3-CNRS et Universit\'e Claude Bernard\\
43, Boulevard du 11 Novembre 1918\\
F-69622 Villeurbanne Cedex, France}

% ======================================
%  You may repeat \author \address as often as necessary      %
% ======================================

\maketitle\abstracts{The Hamiltonian for a fractional supersymmetric
oscillator is derived from three approaches. The first one is based on
the $Q$-uon $\to$ boson $+$ $k$-fermion decomposition. The second one 
starts from a generalized Weyl-Heisenberg algebra. Finally, the third one 
relies on the quantum algebra $U_q(sl_2)$ where $q$ is a root of unity.}   

\section{Generalities about supersymmetry}
 
What is supersymmetry~? Roughly speaking SUperSYmmetry or SUSY
can be defined as a symmetry between bosons and fermions (as considered 
as elementary particles or simply as degrees of freedom). In other words,
SUSY is based on the postulated existence of operators $Q_{\alpha}$ which
transform a bosonic field into a fermionic field and vice versa.  
In  the  context  of  quantum  mechanics, such symmetry operators induce 
a $Z_2$-grading of the Hilbert space of quantum states. In a more 
general way, fractional SUSY corresponds to a $Z_k$-grading for which the
Hilbert
space involves both bosonic degrees of freedom (associated to bosons) and
$k$-fermionic degrees of freedom (associated to $k$-fermions to be
described below) with $k \in {\bf N} \setminus \{ 0 , 1 \}$~; the case
$k=2$ corresponds to ordinary SUSY.

Where does SUSY come from~? The concept of SUSY goes back to the sixtees when
many attemps were done in order to unify {\em external symmetries} (described
by
the Poincar\'e group) and {\em internal symmetries} (described by gauge groups)
{\em for elementary particles}. These attempts led to a no-go theorem by
Coleman and
Mandula~$^{1}$ in 1967 which states that, under reasonable assumptions
concerning the 
$S$-matrix, the unification of internal and external symmetries can be achieved
solely 
through the introduction of the direct product of the Poincar\'e group with the
relevant
gauge group. In conclusion, this unification brings nothing new since it simply
amounts to
consider separately the two kinds of symmetries. A way to escape this no-go
theorem was
proposed by Haag, Lopuszanski and Sohnius~$^{2}$ in 1975: The remedy consists
in replacing
the Poincar\'e group by a Poincar\'e {\em supergroup} (or $Z_2$-{\em graded} 
Poincar\'e {\em group}). 

Let us briefly discuss how to introduce the    Poincar\'e   supergroup. We know
that the
Poincar\'e group has ten generators (six $M_{\mu \nu} \equiv - M_{\nu \mu}$ and
four
$P_{\mu}$ with $\mu,\nu \in \{ 0,1,2,3 \}$): Three $M_{\mu \nu}$ describe
ordinary 
rotations, three $M_{\mu \nu}$ describe special Lorentz 
transformations and the four $P_{\mu}$ describe space-time translations. The
Lie algebra
of the Poincar\'e group is characterized by the commutation relations  
$$
[M , M] = M, \quad [M , P] = P, \quad [P , P] = 0
$$
(written in a symbolic way). The minimal extension of the Poincar\'e group into
a  
Poincar\'e   supergroup gives rise to a Lie {\em superalgebra} (or $Z_2$-{\em 
graded} Lie {\em algebra}) involving the ten generators $M_{\mu \nu}$  and 
$P_{\mu}$ plus four
new generators $Q_{\alpha}$ (with $\alpha \in \{ 1,2,3,4 \}$) referred to as
supercharges.
The Lie superalgebra of the Poincar\'e supergroup is then described
by the commutation relations
$$
[M , M] = M, \quad [M , P] = P, \quad [P , P] = 0
$$
$$
[Q , M] = Q, \quad [Q , P] = 0
$$
and the anticommutation relations
$$
\{ Q , Q \} = 0
$$
(in a symbolic way).

What is the consequence of this increase of symmetries 
(i.e., passing from 10 to 14 generators)
and of the introduction of anticommutators~? In general, an increase of
symmetry yields an
increase of degeneracies. For instance, in condensed matter physics, passing
from    
the tetragonal symmetry to the cubical symmetry leads, from a situation where
the degrees
of degeneracy are 1 and 2, to a situation where the degrees of degeneracy are
1, 2 and 3
(in the absence of accidental degeneracies). Another possible consequence of
the
increase of symmetry is the occurrence of new states or new particles. For
instance,
the number of particles is doubled when 
going from the Schr\"odinger equation (with Galilean invariance) to the Dirac
equation
(with Lorentz invariance): To each particle of mass $m$, spin $S$ and charge
$e$ is
associated an antiparticle of mass $m$, spin $S$ and charge $-e$. In a similar
way,
when passing from the  Poincar\'e group to the  Poincar\'e   supergroup, we
associate 
to a known particle of mass $m$, spin $S$ and charge $e$ a new particle of mass
$m'$,
spin $S' = \left| S \pm \frac{1}{2} \right|$ and charge $e$. (We have $m' \not=
m$ because
supersymmetry is a broken symmetry.) The new particle is called a sparticle or
a
particlino according to whether as the known particle is a fermion or a boson.
In terms
of field theory, the sparticle or particlino field results from the action of a
supercharge $Q_{\alpha}$ on the particle field (and vice versa):
$$
Q_{\alpha} : {\rm fermion} \mapsto {\rm sfermion} = {\rm boson} 
$$
$$
Q_{\alpha} : {\rm boson}   \mapsto {\rm bosino}   = {\rm fermion} 
$$
(see also Ref.~3). We thus speak of a selectron (a particle of spin $0$
associated to the electron) and 
              of a photino   (a particle of spin $\frac{1}{2}$ 
                                                     associated to the photon).

It is not our intention to further discuss SUSY from the viewpoint of
elementary
particles. We shall rather focus our attention on 
some aspects of fractional SUSY and, more precisely, 
on  {\em fractional supersymmetric}  or  {\em para-supersymmetric} 
(one-dimensional) nonrelativistic quantum mechanics,
in the vein of the works in Refs.~4-23. 
From   the point of view of quantum mechanics, a SUSY Hamiltonian
comprises both bosonic and fermionic degrees of freedom. Then, the problem is
to
associate a SUSY Hamiltonian to the Hamiltonian of a given dynamical system.
Here, we 
are interested in  one of the simplest dynamical systems, namely, the harmonic
oscillator. We know that the ordinary oscillator can be described in terms of
boson
operators.  We shall see that the SUSY oscillator can be described in terms of 
boson
and fermion operators (reflecting a $Z_2$-grading) and, more generally, 
that the fractional SUSY oscillator 
can be described in terms of  boson
and $k$-fermion operators (corresponding to a $Z_k$-grading with 
$k \in {\bf N} \setminus \{ 0 , 1 \}$). The rest of this paper is thus devoted
to the
construction (via three methods) of the Hamiltonian for a fractional
supersymmetric 
(or SUSY) oscillator.   

This school was dedicated to the memory of Louis Michel. One of the authors
(M.R.~K.) has had the chance to meet Louis Michel at the {\em Universit\'e 
de Montr\'eal} in 1976. Louis will remain an example for him and for many 
of us. We will all remember the exceptional qualities of the man as a
scientist, 
as a private person, as a public figure and as a teacher.

\section{A quon approach to the SUSY oscillator}

\subsection{The $Q$-uon $\to$ boson $+$ $k$-fermion decomposition}
We shall limit ourselves to give an outline of this decomposition (see
Dunne {\em et al.}~$^{24}$ and Mansour {\em et al.}~$^{25}$ for a more
rigorous mathematical presentation based on the isomorphism between the
braided $Z$-line and the $(z , \theta)$-superspace). We start from a 
$Q$-uon algebra spanned by three operators $a_-$, $a_+$ and $N$ 
satisfying the relationships~$^{26}$
$$
a_- a_+ - Q a_+ a_- = 1
$$
$$
Na_-  -  a_- N = - a_-, \quad Na_+  -  a_+ N = + a_+
$$
where $Q$ is generic (a real number different from zero). Let us consider 
the $Q$-deformed coherent state~$^{27}$
$$
| Z ) := \sum_{n=0}^{\infty} 
\frac{ \left( Z a_+ \right)^n }{ [n]_Q! } \> | 0 \rangle
       = \sum_{n=0}^{\infty} 
\frac{ Z^n }{ ([n]_Q!)^{\frac{1}{2}} } \> | n \rangle
% \eqno (1)
$$ 
\noindent (with $| n \rangle$ such that
$N | n \rangle = n | n \rangle$ and $Z \in {\bf C}$) where 
$$
[n]_Q := {1 - Q^n \over 1 - Q} 
$$
\noindent and
$$
\lbrack n \rbrack_Q ! := 
\lbrack 1 \rbrack_Q \>
\lbrack 2 \rbrack_Q \> \cdots \> 
\lbrack n \rbrack_Q \quad {\hbox{for}} \quad 
n \in {\bf N}^* \quad {\hbox{and}} \quad
\lbrack 0 \rbrack_Q ! := 1
$$
\noindent If we assume that 
$$
Q \to q := \exp \left( \frac{2 \pi {\rm i}}{k} \right), \quad 
k \in {\bf N} \setminus \{ 0,1 \}
$$
then we have $\lbrack k \rbrack_Q ! \to \lbrack k \rbrack_q ! = 0$. Therefore,
in order to give a sense to $| Z )$ for $Q \to q$, we have to do the
replacement
$$
a_+ \leadsto f_+ \quad {\rm with} \quad (f_+)^k = 0
$$
\noindent We thus end up with what we call a $k$-fermionic algebra
spanned by the operators  $f_-$, $f_+$ and $N$ completed by the adjoints
$f_+^+$ and $f_-^+$ of $f_+$ and $f_-$, respectively.~$^{20-22}$ The defining
relations for the
$k$-fermionic algebra are
$$
f_- f_+ - q f_+ f_- = 1
$$
$$
Nf_-  -  f_- N = - f_-, \quad Nf_+  -  f_+ N = + f_+
$$
$$
\left( f_- \right)^k = 
\left( f_+ \right)^k = 0
$$
and similar relations for $f_+^+$ and $f_-^+$. The case $k=2$ corresponds
to ordinary fermion operators and the case $k \to \infty$ 
to ordinary boson   operators. The $k$-fermions are objects interpolating
between
fermions and bosons. They share some properties with the para-fermions~$^{4-6}$
and the anyons as introduced by Goldin {\em et al.}~$^{28}$ (see also Ref.~29).
If we define 
$$
b_{\pm} := \lim_{Q \to q} \frac{ \left( a_{\pm} \right)^k}{[k]_Q !} 
$$
we obtain 
$$
b_- b_+ - b_+ b_- = 1
$$
so that the operators $b_-$ and $b_+$ can be considered as ordinary boson
operators. 
This is at the root of the two following results.~$^{20}$

As a first result, the set $\{ a_- , a_+ \}$ gives rise, for $Q \to q$, 
to two commuting sets: 
The set $\{ b_- , b_+ \}$ of boson operators and the set of $k$-fermion
operators $\{ f_- , f_+ \}$. As a second result, this decomposition leads 
to the replacement of the $Q$-deformed coherent state $ | Z ) $ by the 
so-called fractional supercoherent state
$$
| z , {\theta} ) := \sum_{r=0}^{\infty} 
                    \sum_{s=0}^{k-1} 
                    \frac{ \theta^s }{ ([s]_q!)^{\frac{1}{2}} }
                    \> \frac{ z^r }{ \sqrt{r!} }
                    \> | kr + s \rangle
$$
\noindent where $z$ is a (bosonic) complex variable and $\theta$ a 
($k$-fermionic) generalized Grassmann variable$^{4,6,30,31}$ with $(\theta)^k =
0$.  
The fractional supercoherent state $| z^k , {\theta} )$ can be seen to 
be a linear combination of the coherent states introduced by Vourdas~$^{32}$
with coefficients in the generalized Grassmann algebra spanned by $\theta$ and
the derivative
$\partial_{\theta}$. 
   
In the case $k=2$, the fractional supercoherent state $| z , {\theta} )$
turns out to be a coherent state for the ordinary (or $Z_2$-graded) 
supersymmetric oscillator.~$^{33}$
We construct below a fractional (or $Z_k$-graded) supersymmetric oscillator
which can be associated to the fractional supercoherent state $| z , \theta)$ 
with $k = 3, 4, \cdots$.

\subsection{A quon approach to the Weyl-Heisenberg algebra}
In this section, the basic ingredients consist of a pair of ordinary bosons
$( b_- , b_+ )$ and a pair of $k$-fermions $( f_- , f_+ )$. The $f$'s satisfy
$q$-commutation relations and the $b$'s usual commutation relations 
(see the relations above). In addition, the $f$'s commute with the $b$'s.
Indeed, 
the two pairs $( b_- , b_+ )$ and $( f_- , f_+ )$ may be considered as
originating
from a pair of $Q$-uons $( a_- , a_+ )$ through the above-described
$Q$-{\em uon} $\to$ {\em boson} $+$ $k$-{\em fermion decomposition}. 

Let us define the two operators $X_-$  and   $X_+$   by
$$
X_- := b_- \left[ f_-   +    \frac{(f_+)^{k-1}}{[k - 1]_q !} \right],     
\quad 
X_+ := b_+ \left[ f_-   +    \frac{(f_+)^{k-1}}{[k - 1]_q !} \right]^{k-1}
$$
\noindent and the operator $K$ by
$$
K := f_- f_+ - f_+ f_-
$$
in terms of the operators $b_-$, $b_+$, $f_-$ and $f_+$. Then, 
we can check that $X_-$, $X_+$ and $K$ satisfy
$$
X_- X_+ - X_+ X_- = 1,    \quad K^k = 1
$$
$$
K X_+ - q      X_+ K = 0, \quad 
K X_- - q^{-1} X_- K = 0
$$
\noindent Furthermore, the operator 
$$
M := X_+ X_-
$$ 
\noindent satisfies the following commutation relations
$$
M X_-   -   X_- M   = - X_-,  \quad
M X_+   -   X_+ M   = + X_+,  \quad M K - K M = 0
$$
\noindent The operators $X_-$, $X_+$, $K$ and $M$ thus generate an extended
Weyl-Heisenberg algebra.  

\subsection{The resulting fractional supersymmetric oscillator}
In the spirit of the work by Rubakov and Spiridonov,~$^{4}$ we 
introduce the $k$ projection operators
$$
\Pi_{i} := \frac{1}{k}  \>  \sum_{s=0}^{k-1}  \>  q^{si} \> K^s, \quad i = 0,
1,
\cdots, k-1
$$
\noindent for the cyclic group $Z_k$, the two supercharges
$$
Q_- := X_- (1 - \Pi_{k-1}), \quad  Q_+ := X_+ (1 - \Pi_{0})
$$
\noindent and the Hamiltonian $H$ defined through 
$$
\left( Q_- \right)^{k-1} Q_+   +       \left( Q_- \right)^{k-2} Q_+ Q_- 
                               +   \cdots 
                               +   Q_+ \left( Q_- \right)^{k-1}
                               =       \left( Q_- \right)^{k-2} H
$$
It is then a simple matter of calculation to show that the supercharges
$Q_-$ and $Q_+$ are nilpotent operators with 
$$
\left( Q_- \right)^k = \left( Q_+ \right)^k = 0 
$$
\noindent and that the Hamiltonian $H$ can be written as a bilinear form 
involving $X_- X_+$ and $\Pi_0, \Pi_1, \cdots, \Pi_{k-1}$ with 
\begin{eqnarray*}
H = X_- X_+ \Pi_1 &+& \sum_{\ell = 2}^{k-1}      
                      (X_+ X_- - \ell + 1) (\Pi_0 + \Pi_1 + \cdots + \Pi_{k -
                      \ell - 1 })   \\
                  &+& \sum_{\ell = 2}^{k-1} \ell (X_- X_+ + \frac{\ell - 1}{2})
                  \Pi_{\ell}
                                                + X_+ X_- (1 - \Pi_{k-1})
\end{eqnarray*}
\noindent In addition, we have the commutation relations 
$$
H Q_{\pm} - Q_{\pm} H = 0
$$
and  thus  $Q_{-}$ and $Q_{+}$ can be regarded as constants of motion. Finally,
it can be seen that $H$ may be associated to the fractional supercoherent state 
$| z , \theta )$ with $k \in {\bf N} \setminus \{ 0,1 \}$.  

\subsection{Examples}
\noindent {Example 1.} As a first example, we take $k=2$, i.e., $q=-1$. Then,
we have
$$
X_- := b_- \left( f_-   +    f_+ \right), \quad 
X_+ := b_+ \left( f_-   +    f_+ \right)
$$
\noindent and
$$
K := f_- f_+ - f_+ f_-
$$
\noindent where $( b_- , b_+ )$ are ordinary bosons 
and $( f_- , f_+ )$ ordinary fermions. The operators 
$X_-$, $X_+$ and $K$ satisfy
$$
X_- X_+ - X_+ X_- = 1, \quad K^2 = 1
$$
$$
K X_+   +   X_+ K = 0, \quad 
K X_-   +   X_- K = 0
$$
\noindent which reflect bosonic  and  fermionic 
degrees of freedom. The projection operators
$$
\Pi_0 := \frac{1}{2} (1 + K), \quad 
\Pi_1 := \frac{1}{2} (1 - K)
$$
\noindent are here simple chirality operators and the supercharges
$$
Q_- := X_- \Pi_0, \quad 
Q_+ := X_+ \Pi_1
$$
\noindent have the property
$$
\left( Q_- \right)^2 = \left( Q_+ \right)^2 = 0 
$$
\noindent The Hamiltonian $H$ assumes the form
$$
H = Q_- Q_+  +   Q_+ Q_-
$$
\noindent which can be rewritten as
$$
H = X_+ X_- \Pi_0   +   X_- X_+ \Pi_1 
$$
\noindent It is clear that $H$ commutes with $ Q_- $ 
and   $Q_+$.  In terms of boson and fermion operators, we have
$$
Q_- = f_+ b_-, \quad 
Q_+ = f_- b_+
$$
\noindent and
$$
H = b_+ b_-    +   f_+ f_-
$$
\noindent so that $H$ corresponds to the ordinary (or $Z_2$-graded) 
supersymmetric oscillator whose energy spectrum $E$ is (in a symbolic way)
$$
E = 1 \oplus 2 \oplus 2 \oplus \cdots
$$ 
\noindent with equally spaced levels, the ground state  being  a 
singlet (denoted by 1) and all the excited states (viz., an infinite sequence) 
being doublets
(denoted by 2). Finally, note that the fractional supercoherent state
$ | z , \theta ) $ with $k=2$ is a coherent state for the
Hamiltonian $H$ (see Ref.~33). 

\noindent {Example 2.} We continue with $k=3$, i.e., 
$$
q = \exp \left( \frac{2 \pi {\rm i}}{3} \right)
$$
In this case, we take
$$
X_- := b_- \left[ f_-   +    \frac{(f_+)^2}{[2]_q!} \right],   \quad 
X_+ := b_+ \left[ f_-   +    \frac{(f_+)^2}{[2]_q!} \right]^2
$$
\noindent and
$$
K := f_- f_+ - f_+ f_-
$$
\noindent where $( b_- , b_+ )$ are ordinary bosons 
and $( f_- , f_+ )$ are  $3$-fermions. We hence have
$$
X_- X_+ - X_+ X_- = 1, \quad K^3 = 1
$$
$$
K X_+   -          q        X_+ K = 0, \quad 
K X_-   -          q^{-1}   X_- K = 0
$$
\noindent Our general definitions can be specialized to
\begin{eqnarray*}
\Pi_0 &:=& \frac{1}{3} (1 + q^3 K + q^3 K^2) \\
\Pi_1 &:=& \frac{1}{3} (1 + q^1 K + q^2 K^2) \\
\Pi_2 &:=& \frac{1}{3} (1 + q^2 K + q^1 K^2)
\end{eqnarray*}
\noindent for the projection operators and to  
$$
Q_- := X_- (\Pi_0  + \Pi_1), \quad 
Q_+ := X_+ (\Pi_1  + \Pi_2) 
$$ 
\noindent for the supercharges with the property
$$
\left( Q_- \right)^3 = \left( Q_+ \right)^3 = 0 
$$
\noindent By introducing the Hamiltonian $H$ via
$$
\left( Q_- \right)^{2} Q_+     +   Q_- Q_+ Q_- 
                               +   Q_+ \left( Q_- \right)^{2}
                               =   Q_- H
$$
\noindent we obtain
$$
H = \left( 2 X_+ X_-  -  1 \right) \Pi_0 +
    \left( 2 X_+ X_-  +  1 \right) \Pi_1 +
    \left( 2 X_+ X_-  +  3 \right) \Pi_2
$$
\noindent which can be rewritten as 
$$
H = 2 b_+ b_-  -  1  +  2(1 - 2q) f_+ f_-  +  2(1 + 2 q) f_+ f_- f_+ f_-
$$
\noindent in terms of boson and 3-fermion operators.
We can check that $H$ commutes with $ Q_- $ 
and   $Q_+$. The energy spectrum of $H$ reads 
$$
E = 1 \oplus 2 \oplus 3 \oplus 3 \oplus \cdots
$$ 
\noindent It contains equally spaced levels with a 
nondegenerate ground state (denoted as 1), a doubly 
degenerate first excited state (denoted as 2) and an infinite
sequence of triply degenerate excited states (denoted as 3). 
  
\section{A generalized Weyl-Heisenberg approach to the SUSY oscillator}

This second approach starts directly from a generalized Weyl-Heisenberg 
algebra. Let this algebra be spanned by four operators 
$Y_-$, $Y_+$, $N$ and $K$ (the analogues of 
$X_-$, $X_+$, $M$ and $K$ in Section 2) with the relations 
$$
Y_- Y_+ - Y_+ Y_- = \sum_{s=0}^{k-1} f_s(N) \> \Pi_s, \quad K^k = 1
$$
$$
K Y_+ -          q      Y_+ K = 0, \quad 
K Y_- -          q^{-1} Y_- K = 0
$$
\noindent and 
$$
N Y_-   -   Y_- N   = - Y_-,  \quad
N Y_+   -   Y_+ N   = + Y_+,  \quad  N K - K N = 0
$$
\noindent where $f_s$ is an arbitrary (reasonable) function, $\Pi_s$ the 
projection operator defined in Section 2 and $q$ is the same root of 
unity as before. By defining the supercharges $Q_-$ and $Q_+$ and the
Hamiltonian $H$ as in Section 2 (with $X_{\pm}$ replaced by $Y_{\pm}$), 
we obtain again that  $Q_-$ and $Q_+$
are nilpotent operators of order $k$ which commute with $H$. Furthermore, 
we find that the operator $H$ can be expressed as 
\begin{eqnarray*}
H   &=& \left[ (k-1) Y_+ Y_-  -            \sum_{\ell = 1}^{k-2}  (k-1- \ell)
\> f_{\ell}(N        -1) \right]  \Pi_0 \\
    &+& \left[ (k-1) Y_- Y_+  -            \sum_{\ell = 1}^{k-2}  (k-1- \ell)
\> f_{\ell}(N - \ell +1) \right]  \Pi_1 \\
    &+& \left[ (k-1) Y_- Y_+  + f_1(N+1) - \sum_{\ell = 2}^{k-2}  (k-1- \ell)
\> f_{\ell}(N - \ell +2) \right]  \Pi_2 \\
    &+& \sum_{\ell = 3}^{k-1} 
        \left[ (k-1) Y_+ Y_-             + \sum_{j    = 1}^{\ell} j          
\> f_{j}   (N + \ell -j) \right]  \Pi_{\ell}
\end{eqnarray*}
which generalizes the Hamiltonian derived in Section 2.  

\section{An $U_q(sl_2)$ approach to the SUSY oscillator}

In this section,  we shall briefly show how it is possible to define an
Hamiltonian for a  fractional supersymmetric oscillator in terms of the 
generators of the quantum algebra $U_q(sl_2)$ with $q$ a root of unity.
Let  $J_-$,  $J_+$,  $q^{J_3}$  and  $q^{-J_3}$  be  the  generators of 
$U_q(sl_2)$. They satisfy the relationships 
$$
J_+ J_-  -  J_- J_+  =  \frac{ q^{2J_3}  -  q^{-2J_3} }{ q - q^{-1} }
$$ 
$$
q^{J_3} J_+ q^{-J_3} = q      J_+, \quad 
q^{J_3} J_- q^{-J_3} = q^{-1} J_-
$$
where $q$ is the same root of unity as before. It is straightforward to 
prove that the operator
$$
C := J_- J_+  +  \frac{q q^{2J_3} + q^{-1} q^{-2J_3}} {(q - q^{-1})^2}
$$ 
is an invariant of $U_q(sl_2)$. 

We now define the operator $K$ by
$$
K := q^{J_3}
$$
and the projection operators $\Pi_s$ ($s = 1, 2, \cdots, k-1$) 
as functions of $K$ as in Section 2. The two supercharges  
$Q_-$  and  $Q_+$  are defined here by 
$$
Q_- := J_- (1 - \Pi_{k-1}), \quad  Q_+ := J_+ (1 - \Pi_{0})
$$
\noindent and the Hamiltonian $H$ is defined from the multilinear
form involving $Q_-$  and  $Q_+$ given in Section 2.  As a final result, 
we obtain for $H$ the expression 
\begin{eqnarray*}
H   &=& J_- J_+ \Pi_1 + J_+ J_- ( 1 - \Pi_{k-1} ) \\
    &+& \sum_{m=2}^{k-1} 
    \left[ J_+ J_-  +  ( G(J_3-m+1)  +  G(J_3-m+2)  +  \cdots  +  G(J_3-1) )
\right] \\
    &\times& (\Pi_0 + \Pi_1 + \cdots + \Pi_{k-m-1}) \\
    &+& \sum_{m=2}^{k-1} 
    [ m J_- J_+  -  ( (m-1) G(J_3+1)  +  (m-2) G(J_3+2)  + \cdots \\
    &+&  G(J_3+m-1) ) ] \> \Pi_m
\end{eqnarray*}
where $G$ is defined by
$$
G(X) := \frac{q^{2X} - q^{-2X}}{q - q^{-1}}
$$
The Hamiltonian $H$ is thus given by a formula involving $J_3$ (through the
function $G$)  and  the  product  $J_+ J_-$ (or alternatively the operators 
$J_3$ and $C$). 

\section{Concluding remarks}

We have derived three different Hamiltonians for
a supersymmetric oscillator. At this point, it is worthwhile 
to introduce some links between the three derivations.

In the three derivations, the $Z_k$-grading is introduced via 
the same relation, i.e., $K^k = 1$, although
the introduction of $K$ is specific to each approach. The operator $K$
is: (i) a bilinear form of $k$-fermion operators in the quon approach (Section
2)
originally developed in Ref.~22, (ii) a formal operator in the 
generalized Weyl-Heisenberg approach (Section 3), and (iii) a 
generator of the quantum algebra $U_q(sl_2)$ in the $U_q(sl_2)$ 
approach (Section 4). The $Z_k$-grading manifests itself in Section 2
through the introduction of bosonic degrees of freedom and 
$k$-fermionic degrees of freedom but this is not the case in Sections
3 and 4. 

The connection between the approaches in Sections 2 and 3 is easy to describe: 
To pass from the generalized Weyl-Heisenberg approach to the quon approach, it 
is sufficient to take $f_s = 1$ for $s = 1, 2, \cdots, k-1$. It is also to be 
noted that the generalized Weyl-Heisenberg approach covers some published
works:
(i) for $k=2$, our generalized Weyl-Heisenberg 
approach may be specialized to the cases where 
$$
\sum_{s=0}^{1} f_s(N) \> \Pi_s = 1
$$ 
and 
$$
\sum_{s=0}^{1} f_s(N) \> \Pi_s = 1 + c K, \quad c \in {\bf C}
$$
treated by Plyushchay~$^{15}$ and (ii) for
$k \in {\bf N} \setminus \{ 0 , 1 \}$, by taking 
$$
\sum_{s=0}^{k-1} f_s(N) \> \Pi_s = 
\sum_{s=0}^{k-1} c_s K^s, \quad  c_s \in {\bf C}
$$
our generalized Weyl-Heisenberg approach may be 
found to be equivalent to the one developed by 
Quesne and Vansteenquite.~$^{19}$

Finally, note that the connection between the approach 
in Section 4 and the ones in Sections 2 and 3 still 
deserves to be explored. This will be the object of a forthcoming paper. 

To close this paper,  let  us  mention  three domains where
(fractional)  supersymmetry play or might play an important
r\^ole.  First,  in condensed matter physics,  the interest
of fractional supersymmetry rests on its application to the 
fractional quantum Hall effect and supraconductivity 
at critical temperature. Second, in nuclear physics,
it seems that (ordinary) supersymmetry has been observed in
nuclear spectroscopy:   Supersymmetry could relate the structure 
of odd-odd nuclei to even-even and odd-$A$ systems.~$^{34}$ 
Third,  in  elementary  particle  physics,  (ordinary) 
supersymmetry imposes some limits on the masss of the Higgs
boson~: The mass should be below 130 GeV/$c^2$ so that the
``events''  observed  at  LEP200  (CERN)  in  August  2000 around  
115 GeV/$c^2$ are compatible with the hypotheses of supersymmetry 
in high energy physics. As a conclusion, supersymmetry 
is still a chapter of physics under construction.

\section*{Acknowledgments}

The senior author (M.R.~K.) is grateful to the 
Organizing Committee of the 6th International School of Theoretical Physics 
SSPCM'2000, and more specifically Prof.~T.~Lulek, for inviting 
him to deliver this lecture. He is also indebted to Profs.~M.~Bo\.zejko, 
J.D.~Louck and A.~Vourdas for interesting comments.

\end{document}